# Higher-Order Concurrency for Microcontrollers

Abhiroop Sarkar
Robert Krook
Bo Joel Svensson
Mary Sheeran
Chalmers University
Gothenburg, Sweden
{sarkara,krookr,joels,mary.sheeran}@chalmers.se

## Abstract

Programming microcontrollers involves low level interfacing with hardware and peripherals that are concurrent and reactive. Such programs are typically written in a mixture of C and assembly using concurrent language extensions (like `FreeRTOS tasks` and `semaphores`), resulting in unsafe, callback-driven, error-prone and difficult-to-maintain code.

We address this challenge by introducing SenseVM - a bytecode-interpreted virtual machine that provides a message passing based *higher-order concurrency* model, originally introduced by Reppy, for microcontroller programming. This model treats synchronous operations as first-class values (called `Events`) akin to the treatment of first-class functions in functional languages. This primarily allows the programmer to compose and tailor their own concurrency abstractions and, additionally, abstracts away unsafe memory operations, common in shared-memory concurrency models, thereby making microcontroller programs safer, composable and easier-to-maintain.

Our VM is made portable via a low-level *bridge* interface, built atop the embedded OS - Zephyr. The bridge is implemented by all drivers and designed such that programming in response to a software message or a hardware interrupt remains uniform and indistinguishable. In this paper we demonstrate the features of our VM through an example, written in a Caml-like functional language, running on the `nRF52840` and `STM32F4` microcontrollers.

## 1 Introduction

Microcontrollers are ubiquitous in embedded systems and IoT applications. These applications, like robot controllers, cars, industrial machinery, are inherently concurrent and event-driven (termed *reactive* in more recent literature). A 2011 poll[26] conducted among embedded systems developers found C, C++ and assembly to be the most popular choices for programming such reactive applications.

The popularity of the C family can be attributed to its fine-grained control over memory layout and management. Also, C compilers are extremely portable across a diverse set of microcontrollers while offering low-level control over hardware peripherals. However, C not being an innately concurrent language, embedded OSes like FreeRTOS and Zephyr provide their own shared memory concurrency abstractions like threads, semaphores, mutexes etc [32–34]. Additionally, the event-driven driver interfaces in such OSes tend to have APIs that look like the following:

```c
int gpio_pin_interrupt_configure(const struct device *port
                                , gpio_pin_t pin
                                , gpio_flags_t flags);
void gpio_init_callback(struct gpio_callback *callback
                      , gpio_callback_handler_t handler
                      , gpio_port_pins_t pin_mask);
int gpio_add_callback(const struct device *port
                    , struct gpio_callback *callback);
```

This combination of programming in a memory-unsafe language, like C, callback-based driver APIs and shared-memory concurrency primitives leads to error-prone, difficult-to-maintain and unsafe programs. Moreover, programs also end up being very difficult to port to other microcontroller boards and follow intricate locking protocols and complex state machines to deal with the concurrent and reactive nature of the applications.

Recently, there has been a surge in dynamic-language-based runtime environments like MicroPython [9] and Espruino [31] on microcontrollers. While these languages abstract away the unsafe memory management of the C family, neither of them is inherently concurrent. Programming with the hardware peripherals in MicroPython has the following form:

```
def callback(x):
#...callback body with nested callbacks...
extint = pyb.ExtInt(pin, pyb.ExtInt.IRQ_FALLING
                   , pyb.Pin.PULL_UP, callback)
ExtInt.enable()
```

The above is plainly a wrapper over the original C API and is prone to suffer from the additional and *accidental* complexity of nested callback programming, colloquially termed as *callback-hell*[18].

**Contributions.** In this paper we simplify the handling of callback-driven APIs, discussed above, by introducing a bytecode-interpreted virtual machine, SenseVM[1], which models all hardware and I/O interactions via a message-passing interface. We enumerate the practical contributions of SenseVM below:

1. **Higher-order concurrency**. We provide, to the best of our knowledge, the first implementation of the higher-order concurrency model [20] for programming microcontrollers. This model allows the introduction of first class values, called Events, for representing synchronous operations and provides combinators to compose more complex Event trees, which are useful for control-flow heavy programs, common in microcontrollers. We briefly summarize the model in Section 2.1 and describe implementation details in Section 3.
2. **Message-passing based I/O**. Noting that complex state machines and callback-hell arises from the intertwined nature of *concurrency* and *callback-based I/O*, we mitigate the issues by unifying concurrency with I/O. As a result, programming in response to a hardware interrupt or any other I/O message in SenseVM remains indistinguishable from responding to a software message. Moreover, owing to the higher-order concurrency model, the programs do not reduce to a chain of *switch-casing* of message contents, common in other message passing languages like Erlang. We explain implementation details of the message passing (Section 3.2), show a sample program (Section 2.3) and evaluate the performance metrics of this I/O model (Section 4).
3. **Portability**. Portability among microcontroller boards is challenging, as a consequence of the diverse set of peripherals and hardware interfaces available. To address this, SenseVM provides a *low-level bridge* interface written in C99 (described in Section 3.3) which provides a common API which can be implemented by various synchronous and asynchronous drivers. Once drivers of different boards implement this interface, programs can be trivially ported between them as we demonstrate by running the same, unaltered program on the nRF52840 and STM32F4 based boards.

## 2 Programming on SenseVM

To demonstrate programming on top of the SenseVM, we will use an eagerly-evaluated, statically-typed, functional language which extends the polymorphic lambda calculus with let and letrec expressions similar to Caml [16]. Unlike Caml, it lacks the mutable ref type and does all I/O operations via the message-passing interface of the VM. Type declarations and signatures in our language are syntactically similar to those of Haskell [14].

The message-passing interface of SenseVM is exposed via runtime supplied functions. It is *synchronous* in nature and all communications happen over *typed channels*. In a Haskell-like notation, the general type signature of message sending and receiving over channels could be written:

```
sendMsg : Channel a -> a -> ()
recvMsg : Channel a -> a
-- a  denotes any generic type like Int, Bool etc;
-- () denotes the "void" type
```

However, our VM implements an extension of the above known as *higher-order concurrency* [20]. We describe the differences and their implications on the programming model in the following section.

### 2.1 Higher-Order Concurrency

The central idea of *higher-order concurrency* is to separate the act of synchronous communication into two steps -

1. Expressing the intent of communication
2. Synchronisation between the sender and the receiver

The first step above produces first-class values called Events, which are concrete runtime values, provided by the SenseVM. The second step, synchronisation, is expressed using an operation called sync. Now we can write the type signature of message passing in our VM as:

```
send : Channel a -> a -> Event ()
recv : Channel a -> Event a
sync : Event    a -> a
```

Intuitively, we can draw an equivalence between general message passing and higher-order concurrency based message passing, using function composition, like the following:

```
sync . send ≡ sendMsg
sync . recv ≡ recvMsg
```

---

[1]https://github.com/svenssonjoel/Sense-VM

The above treatment of "Events as values" is analogous to the treatment of "functions as values" in functional programming. In a similar spirit as higher-order functions, our VM provides Event based combinators for further composition of trees of Events:

```
choose : Event a -> Event a  -> Event a
wrap   : Event a -> (a -> b) -> Event b
```

The choose operator represents the standard *selective communication* mechanism, necessary for threads to communicate with multiple partners, found in CSP [12]. The wrap combinator is used to run *post-synchronisation* operations. wrap ev f can be read as - "post synchronisation of the event ev, apply the function f to the result".

Reppy draws parallels between an Event and its associated combinators with higher-order functions [21], using the following table:

| Property | Function values | Event values |
| --- | --- | --- |
| Type constructor | -> | event |
| Introduction | λ - abstraction | receive |
| | | send |
| | | etc. |
| Elimination | application | sync |
| Combinators | composition | choose |
| | map | wrap |
| | etc. | etc. |

SenseVM provides other combinators like spawn for spawning a *lightweight* thread and channel for creating a typed channel:

```
spawn   : (() -> ()) -> ThreadId
channel : ()  -> Channel a
```

Next we discuss the message-passing API for handling I/O.

### 2.2 I/O in SenseVM

The runtime APIs, introduced in the previous section, are useful for implementing a software message-passing framework. However, to model external hardware events, like interrupts, we introduce another runtime function:

```
spawnExternal : Channel a -> Int -> ThreadId
```

This function connects the various peripherals on a microcontroller board with the running program, via typed channels. The second argument to spawnExternal is a *driver-specific identifier* to identify the driver that we wish to communicate with. Currently in our runtime, we use a monotonically increasing function to number all the drivers starting from 0; the programmer uses this number in spawnExternal. However, as future work, we are building a tool that will parse a configuration file describing the drivers present on a board, number the drivers, inform the VM about the numbering and then generate a frontend program like the following:

```
data Driver = LED Int | Button Int | ...

led0 = LED 0
led1 = LED 1
but0 = Button 2
but1 = Button 3

--Revised `spawnExternal` type signature will be
spawnExternal : Channel a -> Driver -> ThreadId
```

In the rest of the paper, we will be referring to the revised spawnExternal function for clarity. Now we can express a SenseVM program that listens to an interrupt raised by a button press below:

```
main =
  let bchan = channel() in
  let _ = spawnExternal bchan but0 in
  sync (recv bchan)
```

The recv is a *blocking* receive and if there are other processes that can be scheduled while this part of the program is blocked, the runtime will schedule them. In the absence of any other processes, the runtime will relinquish its *control* to the underlying OS (Zephyr OS) and will **never poll** for the button press. The SenseVM runtime is geared towards IoT applications where microcontrollers are predominantly asleep and are woken up reactively by hardware interrupts. In the following section, we demonstrate a more complete program running on our VM.

### 2.3 Button-Blinky

Next we *portably* run a program in both the nRF52840 and STM32F4 microcontroller based boards. The program *indefinitely* waits for a button press on button 0 and on receiving an ON signal, sends an ON signal to LED number 0. Upon the button release, it receives an OFF signal and sends the same to the LED. The LED stays ON as long as the button is pressed.

This program, expressed in C and hosted in Zephyr, is about 127 lines of code (see [7]) involving setup, initialisation, callback registration and other control logic. The same program expressed on top of the SenseVM is:

```
1   bchan = channel ()
2   lchan = channel ()
3
4   main =
5     let _ = spawnExternal bchan but0 in
6     let _ = spawnExternal lchan led0 in
7     buttonBlinky
8
9   buttonBlinky =
10    let _ = sync (wrap (recv bchan) blinkled) in
11    buttonBlinky
12    where
13      blinkled i = sync (send lchan i)
```

In the above program, we create a channel per driver (Line no. 1 and 2) and instruct the button driver, via spawnExternal, to send any hardware interrupts to the registered channel - bchan (Line no. 5). Upon receiving an interrupt, we run a post-synchronisation action using wrap (Line no. 10), which sends the value sent by the interrupt to the LED driver using lchan (Line no. 13). It recursively calls itself to continue running the program infinitely (Line no. 11). There is a notable absence of callbacks in the above program.

## 3 Design and Implementation

### 3.1 System Overview

SenseVM, including its execution unit, is an implementation of the Categorical Abstract Machine [5] (CAM), as explained by Hinze [11], but has been augmented with a set of operation codes for the higher-order concurrency extensions. We show the compilation and runtime pipeline of the VM in Figure 1 below:

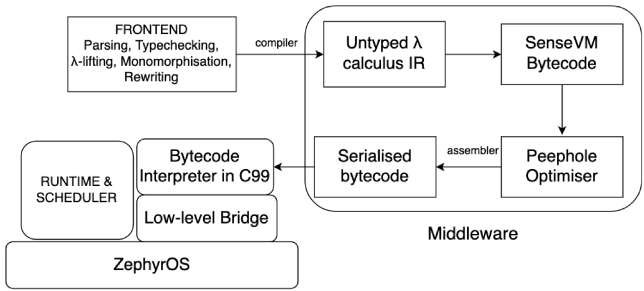

**Figure 1.** The SenseVM compilation & runtime pipeline

*Frontend.* The frontend supports a polymorphic and statically typed functional language whose types are monomorphised at compile time. It supports some optimisation passes like lambda-lifting to reduce heap allocation.

*Middleware.* The frontend's desugared intermediate representation is compiled to an untyped lambda calculus representation. This representation is further optimised by generation of specialised bytecodes for an operational notion called *r-free* variables, as described by Hinze[11], to further reduce heap allocation. The generated SenseVM bytecode operates on a stack machine with a single environment register. The bytecode is then further subjected to peephole optimisations like $\beta$−reduction and last-call optimisation[11] (a generalisation of tail-call elimination).

*Backend.* The SenseVM back-end, or virtual machine, is split into a high-level part and a low-level part. Currently the low-level part is implemented on top of Zephyr and is described in more detail in Section 3.3. The interface between the low-level, Zephyr based, part of the back-end and the high-level has been kept minimal to enable plugging in other embedded OSes like FreeRTOS.

The high-level part of the back-end consists of a fixed number of *contexts*. A context is a lightweight thread comprising of (1) an environment register, (2) a stack and (3) a program counter. The context switching is *cooperatively* handled by the VM scheduler. The high-level part of the VM also contains a garbage-collected heap where all compound CAM values (like tuples) are allocated.

The VM uses a mark-and-sweep garbage collection algorithm that is a combination of the Hughes lazy sweep algorithm and the Deutsch-Schorr-Waite pointer-reversing marking algorithm [13, 15, 23]. As future work, we intend to investigate more static memory management schemes like regions [27] and also real-time GC algorithms[17]. The following section explains the message-passing based concurrency aspects of the VM.

### 3.2 Synchronous Message Passing

In the higher-order concurrency model of synchronous message passing, we separate the *synchronisation* from the *description* of message passing. By doing this, we introduce an intermediate value type known as an Event.

#### 3.2.1 Event.
Reppy calls send and recv operations *base-event* constructors. Operations like choose and wrap are the higher-order operators used for composing the base events. An Event is a concrete runtime value represented in the SenseVM heap as a nested tuple. Figure 2 shows the representation of an Event on the heap.

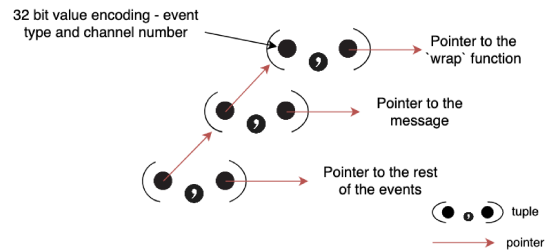

**Figure 2.** The heap structure of an Event

An Event is represented as a linked list of base events. We represent the linked list as a chain of tuples where the second element points to the rest of list, and the first element points to another nested tuple, which encodes - (1) message content, (2) channel number, (3) event type (send or recv) and (4) post-synchronisation function to be applied.

A composition operation like choose e1 e2 simply appends two lists - e1 and e2. We can further compose choose using fold operations[2] to allow it to accept a list of events and build complex event trees, along with wrap, as shown in the following program. However it is always possible to rewrite this tree at compile time, using function composition,

---
[2]https://hackage.haskell.org/package/base-4.15.0.0/docs/Prelude.html#v:foldr1

to produce the canonical representation of events as a linked list rather than a tree.

```
choose' : [Event a] -> Event a
choose' = foldr1 choose

choose' [wrap bev1 w1,
         wrap (choose' [wrap bev2 w2,
                        wrap bev3 w3]) w4]
-- Rewritten to
choose' [ wrap bev1 w1,
          wrap bev2 (w4 . w2),
          wrap bev3 (w4 . w3) ]
```

### 3.2.2 Synchronisation.
The sync or synchronisation operation is one of the more intricate aspects of the SenseVM runtime and we describe the algorithm in pseudocode below:

> **Function** *sync* (*eventList*)
> *ev* ← *findSynchronisableEvent*(*eventList*)
> **if** *ev* ≠ ∅ **then**
> 　*syncNow*(*ev*)
> **else**
> 　*block*(*eventList*)
> 　*dispatchNewThread*()
> **end if**
> **EndFunction**

The above gives a bird's-eye view of the major operations involved in sync. To understand the findSynchronisableEvent function, we should understand the structure of a *channel*, that contains a send and a receive queue. These queues are used not to hold the messages but to track which threads are interested in sending or receiving on the respective channel. Now we can describe findSynchronisableEvent :

> **Function** *findSynchronisableEvent* (*eventList*)
> **for all** *e* ∈ *eventList* **do**
> 　**if** *e.channelNo communicating with driver* **then**
> 　　**if** *llBridge.driver readable/writeable?* **then**
> 　　　**return** *e*
> 　　**end if**
> 　**else**
> 　　**if** *e.baseEventType* == *SEND* **then**
> 　　　**if** ¬*isEmpty*(*e.channelNo.recvq*) **then**
> 　　　　**return** *e*
> 　　　**end if**
> 　　**else if** *e.baseEventType* == *RECV* **then**
> 　　　**if** ¬*isEmpty*(*e.channelNo.sendq*) **then**
> 　　　　**return** *e*
> 　　　**end if**
> 　　**end if**
> 　**end if**
> **end for**
> **return** ∅
> **EndFunction**

When no synchronisable event is found we use block :

> **Function** *block* (*eventList*)
> **for all** *e* ∈ *eventList* **do**
> 　**if** *e.baseEventType* == *SEND* **then**
> 　　*enqueue*(*e.channelNo.sendq*)
> 　**else if** *e.baseEventType* == *RECV* **then**
> 　　*enqueue*(*e.channelNo.recvq*)
> 　**end if**
> **end for**
> **EndFunction**

After the call to block, we call dispatch described below:

> **Function** *dispatchNewThread* ()
> **if** *readyQ* ≠ ∅ **then**
> 　*threadId* ← *dequeue*(*readyQ*)
> 　*currentThread* = *threadId*
> **else**
> 　*relinquish control to Zephyr*
> **end if**
> **EndFunction**

On receiving a synchronisable event, we apply syncNow :

> **Function** *syncNow* (*event*)
> **if** ¬*hardware*(*event*) **then**
> 　**if** *event.baseEventType* == *SEND* **then**
> 　　*threadIdR* ← *dequeue*(*event.channelNo.recvq*)
> 　　*recvEvt* ← *threadIdR.envRegister*
> 　　*event.Message* → *threadIdR.envRegister*
> 　　*threadIdR.programCounter* → *recvEvt.wrapFunc*
> 　　
> 　　*currentThread.programCounter* → *event.wrapFunc*
> 　　
> 　　*sendingThread* = *currentThread*
> 　　*currentThread* = *threadIdR*
> 　　*schedule*(*sendingThread*)
> 　**else if** *event.baseEventType* == *RECV* **then**
> 　　*threadIdS* ← *dequeue*(*event.channelNo.sendq*)
> 　　*sendEvt* ← *threadIdS.envRegister*
> 　　*sendEvent.Message* → *currentThread.envRegister*
> 　　
> 　　*threadIdS.programCounter* → *sendEvt.wrapFunc*
> 　　
> 　　*currentThread.programCounter* → *event.wrapFunc*
> 　　
> 　　*schedule*(*threadIdS*)
> 　**end if**
> **else if** *hardware*(*event*) **then**
> 　**if** *event.baseEventType* == *SEND* **then**
> 　　*llBridge.write*(*event.Message*) → *driver*
> 　**else if** *event.baseEventType* == *RECV* **then**
> 　　*currentThread.envRegister* ← *llBridge.read*(*driver*)
> 　**end if**
> 　*currentThread.programCounter* → *event.wrapFunc*
> **end if**
> **EndFunction**

In the case of interrupts that arrive when the SenseVM runtime has relinquished control to Zephyr, we place the message on the Zephyr queue and request the scheduler to wake up, create an event and then call sync on that event. For events in a choose clause that fail to synchronise, we clear them from the heap using the *dirty-flag* technique invented by Ramsey [19].

### 3.2.3 Comparison with Actors.
In SenseVM, we opt for a synchronous style of message-passing, as distinct from the more popular asynchronous style found in actor-based systems[10]. Our choice is governed by the following:

(1) Asynchronous send (as found in actors) implies the *unboundedness* of an actor mailbox, which is a poor assumption in memory-constrained microcontrollers. With a bounded mailbox, actors eventually resort to synchronous send semantics, as found in SenseVM.

(2) Synchronous message passing can emulate buffering by introducing intermediate processes, which forces the programmer to think upfront about the cost of buffers.

(3) Acknowledgement becomes an additional, explicit step in asynchronous communication, leading to code bloat. Acknowledgement is implicit in the synchronous message passing model and is a practical choice when not communicating across distributed systems.

(4) Actor based systems (like Medusa [3]) incur the extra cost of tagging a message to identify which entities have requested which message. The combination of *channels* and synchronous message-passing ensures that an arriving hardware interrupt knows where to send the message, without any message identification cost.

## 3.3 Low-Level Bridge

The *low-level bridge* exists to bridge the divide in abstraction level from the channel-based communication of SenseVM to the underlying embedded OS - Zephyr [1]. The main motivation behind using the Zephyr RTOS as the lowest level hardware abstraction layer (HAL) for SenseVM is that it provides a good set of driver abstractions that are portable between microcontrollers and development boards. Examples of such drivers include UART, SPI, I2C, buttons and LEDs and many other peripherals.

A driver can be be either synchronous or asynchronous in nature. An example of a synchronous driver is an LED that can be directly read or written. On the other hand, asynchronous peripherals operate using interrupts. An asynchronous driver interrupt can be signaling, for example, that it has filled up a buffer in memory using Direct Memory Access (DMA) that now needs handling, or it could signal a simple boolean message such as "a button has been pressed".

To accommodate these various types of drivers, the connection between SenseVM and the drivers is made using a datatype called ll_driver_t, which describes an interface that each low-level (platform and HAL dependent) driver should implement. The interface supports the following operations:

```
uint32_t ll_read(ll_driver_t *drv,
                 uint8_t *data,
                 uint32_t data_size);
uint32_t ll_write(ll_driver_t *drv,
                  uint8_t *data,
                  uint32_t data_size);
uint32_t ll_data_readable(ll_driver_t *drv);
uint32_t ll_data_writeable(ll_driver_t *drv);
bool ll_is_synchronous(ll_driver_t *drv);
```

For a synchronous driver such as the LED, `ll_read` and `ll_write` can be called at any time to read or write data from the driver. Likewise, the `ll_data_readable` and `ll_data_writeable` always returns a value greater than zero in the synchronous case. The `ll_is_synchronous` function is used by the SenseVM runtime to distinguish asynchronous, interrupt-driven drivers from the synchronous ones and handle them accordingly.

The asynchronous case is more interesting. Say that the program wants to write a value to an asynchronous driver, for example a UART (serial communication). The low-level UART implementation may be using a buffer that can be full and `ll_data_writeable` will in this case return zero. The SenseVM task that tried to write data will now block. Tasks that are blocked on either asynchronous read or write have to be woken up when the lower-level driver implementation has *produced* or *is ready to receive* data. These state changes are usually signaled using interrupts.

The interrupt service routines (ISRs) associated with asynchronous drivers talk to the SenseVM RTS using a message queue and a driver message type called `ll_driver_msg_t` akin to the Medusa system [3]. A simple event such as "a button has been pressed" can be fully represented in the message, but if the message signals that a larger buffer is filled with data that needs processing, this data can be accessed through the `ll_read/write` interface.

The SenseVM scheduler runs within a Zephyr thread. This thread is the owner of the message queue to which all the drivers are sending messages. When the Zephyr thread starts the SenseVM scheduler, it passes along a pointer to a `read_message` function that reads a message from the message queue or blocks if the queue is empty. Thus, all code that relies on Zephyr is confined to the low-level drivers and to the thread that runs the scheduler.

## 4 Evaluation

Evaluations are run on the STM32F4 microcontroller with a Cortex M4 core at 168MHz (STM32F407G-DISC1) and an nRF52840 microcontroller with an 80MHz Cortex M4 core (UBLOX BMD340). We should point out that the button-blinky program runs *portably* on both of these boards with no change in the code required.

### 4.1 Power Consumption

Figure 3 shows the power consumption of the button-blinky program measured in the nRF52 board for (i) a polling implementation of the program written in C, (ii) the interrupt-based implementation written in C [7] and (iii) the SenseVM implementation. The measurements were obtained using a Ruideng UM25C ammeter where the values provided are momentary and read after they stabilised.

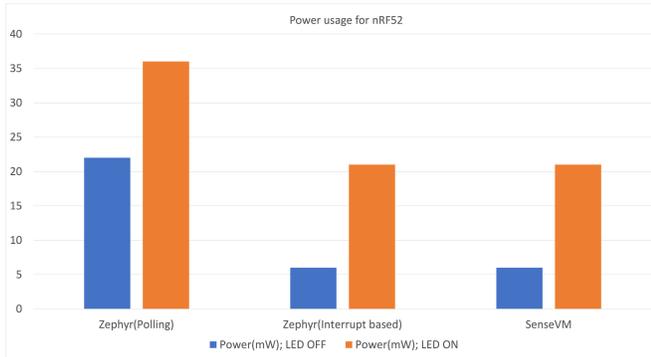

**Figure 3.** Power consumption with LED OFF and ON in mW

The figure shows that the polling implementation consumes up to four times more power when the LED is switched OFF and up to twice as much power when the LED is ON compared to the SenseVM implementation. The C interrupt based implementation and the SenseVM program have identical power usage. However, the C program is 127 lines of callback-based code compared to the 13 line SenseVM program shown in Section 2.3.

### 4.2 Response Time

Table 1 shows two different implementations of button-blinky written directly in C and Zephyr in comparison to the SenseVM implementation. The values in the table represent the time it takes from a button being pressed until the associated LED turns on. The values were obtained using a UNI-T UTD2102CEX (100MHz, 1GS/s oscilloscope) that has not been calibrated other than the original factory calibrations. We elaborate on the oscilloscope measurements in Appendix A

**Table 1.** Response time for button-blinky program in $\mu$secs

|  | Zephyr | Zephyr (Interrupt based) | SenseVM |
|---|---|---|---|
| STM32F4 | 0.88 | 9.5 | 37.7 |
| NRF52 | 1.9 | 17.3 | 61.6 |

The response time for the polling based implementation is considerably faster than the rest. This comes at the cost of being much more power intensive (upto 4 times for this program). The SenseVM response time is up to 4 times slower than the more realistic interrupt based C program. This slowdown is majorly attributed to two factors i) the interpretation overhead and ii) the *stop-the-world* garbage collector. As future work, we plan to experiment with various *ahead-of-time* and *just-in-time* compilation strategies to speed up the bytecode interpretation and thus improve the response times. We also hope that more incremental, real-time garbage collectors can further speed up the response times.

### 4.3 Memory Usage

SenseVM statically allocates the stack, heap, number of channels, number of threads etc and allows configuration of their sizes. As a result, when comparing memory usage between SenseVM and Zephyr, a lot of the memory reported is not currently being used but statically allocated for *potential use* by the VM. We show some memory usage statistics in Table 2.

**Table 2.** Memory usage of button-blinky in KiB (= 1024 bytes)

|  | Flash | RAM |
|---|---|---|
| SenseVM | 37.4 | 27.6 |
| Zephyr message queue | 16.5 | 4.55 |

While obtaining the data in Table 2, SenseVM was configured with a heap of 1024 bytes and 1024 bytes for use by the stacks of each context. Additionally we set aside 9600 bytes for channels, of which, only 2 channels (192 bytes) are used and the remaining 9408 bytes remain unused. The button-blinky program uses 1 thread, 2 channels and interacts with 2 drivers (LED and button), while in the VM we have statically allocated space for 4 threads, 100 channels and metadata for 16 drivers. This additional space allocation is done keeping in mind multi-threaded programs with larger inter-process networks communicating via several channels using various drivers.

## 5 Related Work

Among statically-typed languages, Varoumas et.al [30] presented a virtual machine that can host the OCaml language on PIC microcontrollers. They have extended OCaml with a deterministic model of concurrency - OCaLustre [29]. This line of work, however, does not mitigate the pain associated with *callback hell*, as the concurrency model does not extend to the interrupts and their handlers. In SenseVM, we unify the notion of concurrency and I/O (callback-based and otherwise) via message-passing to simplify callback-oriented programming, prevalent in microcontrollers.

In the dynamically-typed language world there exists Medusa [3], for programming the TI Stellaris microcontrollers, which is much closer to our line of work by unifying concurrency and I/O. The difference between Medusa and SenseVM boils down to the comparison between actor based, asynchronous message passing systems and the synchronous message passing model, which has been discussed in Section 3.2.3. Moreover, the static typing of our frontend language enables *typed channels* to perform static checks on message contents, done at runtime in Medusa. One should note, however, that actor based systems excel at distributed computing - a more failure-prone and harder form of concurrent computing.

Programming environments like VeloxVM [28] focus on the safety and security of microcontroller programming . We leave our investigations on the formal safety and security aspects of microcontroller programming as future work. There also exists work to improve the portability, debuggability and live-code updating capability of microcontrollers using WebAssembly [25].

Our previous work [22] has attempted to use the Functional Reactive Programming (FRP) paradigm [6] for the programming of microcontrollers. However, it suffered from the performance penalties of *polling*, owing to the pull-based semantics of FRP, which has been addressed in this higher-order concurrency model.

The higher-order concurrency model, first introduced by Reppy [20], has been primarily used for programming GUIs such as eXene [8]. It has also found application in more experimental distributed implementations of SML such as DML [4]. We provide, to the best of our knowledge, the first implementation of the model for natively programming microcontrollers. In all previous implementations of higher-order concurrency, external I/O has been represented using stream-based I/O primitives on top of SML's standard I/O API [21]. We differ in this aspect by modelling any external I/O device as a process in itself (connected using spawnExternal) and communicating via the standard message passing interface, making the model more uniform.

Regarding portability, there exist JVM implementations like Jamaica VM [24] for running portable Java code across various classes of embedded systems. The WebAssembly project has also spawned sub-projects like the WASM micro-runtime [2] to allow languages that compile to WebAssembly to run portably on microcontrollers. It should be pointed out that while general-purpose languages like Javascript can execute on ARM architectures by compiling to WebAssembly, it is still the case that neither the language nor the VM offers any intrinsic model of concurrency unified with I/O, analogous to the one provided by SenseVM.

## 6 Conclusion

We have presented SenseVM, a bytecode-interpreted virtual machine to simplify the concurrent and reactive programming of microcontrollers. We introduce the higher-order concurrency model to the world of microcontroller programming and show the feasibility of implementing the model by presenting the power usage, response time measurements and memory footprint of the VM on an interrupt-driven, callback based program. We additionally present our VM to address the portability concerns that plague low-level C and assembly microcontroller programs. We demonstrate the portability of our VM by running the same program, unchanged, on the nRF52840 and STM32F4 microcontrollers. In future work, we hope to experiment with real-time programming on microcontrollers by exploring more deterministic approaches to memory management like regions and real-time garbage collectors.

## Acknowledgments

This work was funded by the Swedish Foundation for Strategic Research (SSF) under the project Octopi (Ref. RIT17-0023) and by the Chalmers Gender Initiative for Excellence (Genie).

## A Response Time Measurement Data

In the pictures below, the blue line represents the button and the yellow line represents the LED. The measurement is taken from the rising edge of the button line, to the rising edge of the LED while trying to (when possible) cut the slope of the edge in the middle. All pictures are captured on the same UNI-T UTD2102CEX (100MHz, 1GS/s oscilloscope).

The oscilloscope used here is not pleasant to gather large datasets. To the best of our knowledge it does not support a network connection and API access.

The occurrence of outliers in our measurements is discussed in the next section. A series of 25 SenseVM button-blinky response time measurements, filtered from outliers, were collected on the STM32F4 microcontroller. Out of these 25 measurements 23 were measured at 37.7$\mu$s and the other two came in at 37.4$\mu$s and 37.5$\mu$s.

## B Outliers

While repeatedly running and capturing the response time of the button-blinky program on the STM32F4, periodical outliers occur. In a sequence of 108 experiments, 11 showed

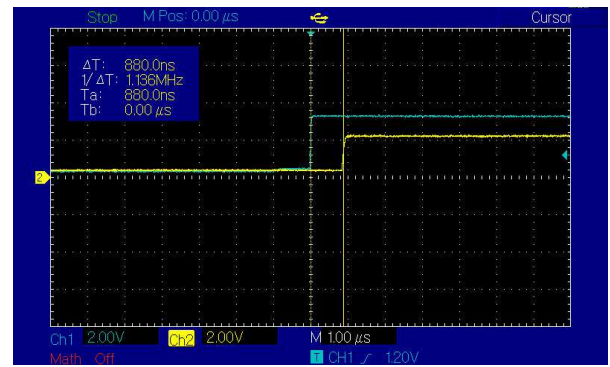

**Figure 4.** STM32F4 Discovery: Zephyr implementation of button-blinky that constantly polls button and updates LED.

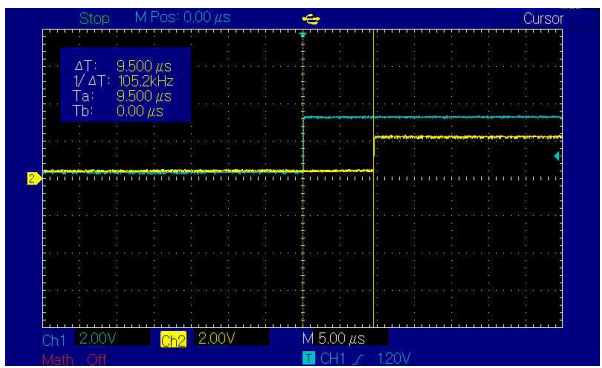

**Figure 5.** STM32F4 Discovery: Zephyr implementation of button-blinky using interrupt and message queue.

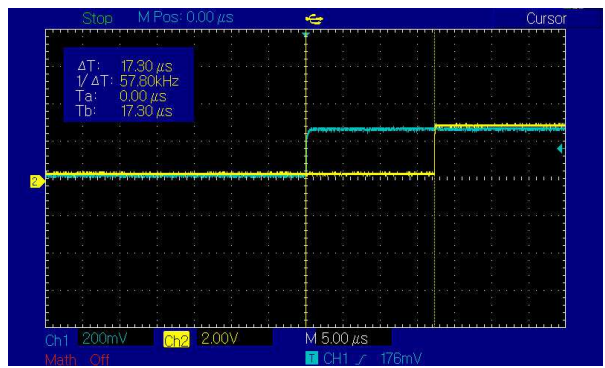

**Figure 8.** NRF52: Zephyr implementation of button-blinky using interrupt and message queue.

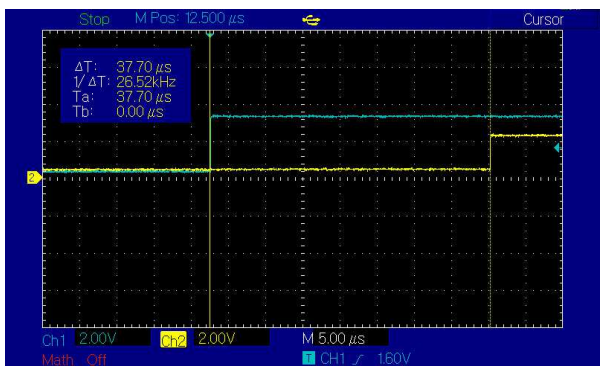

**Figure 6.** STM32F4 Discovery: SenseVM implementation of button-blinky.

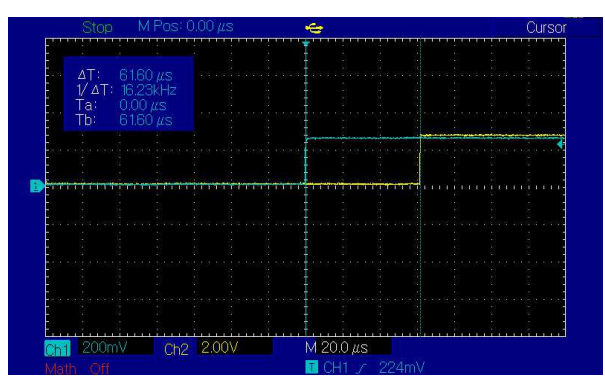

**Figure 9.** NRF52: SenseVM version of button-blinky.

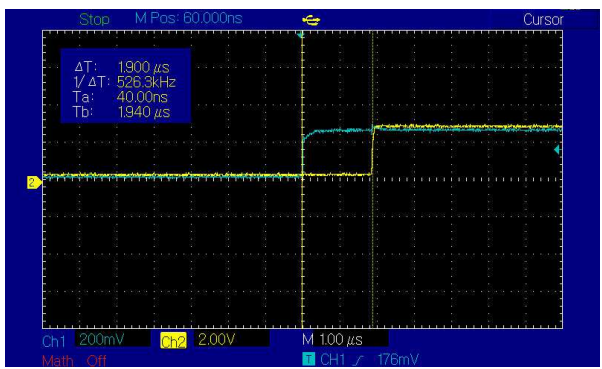

**Figure 7.** NRF52: Zephyr implementation of button-blinky that constantly polls button and updates LED.

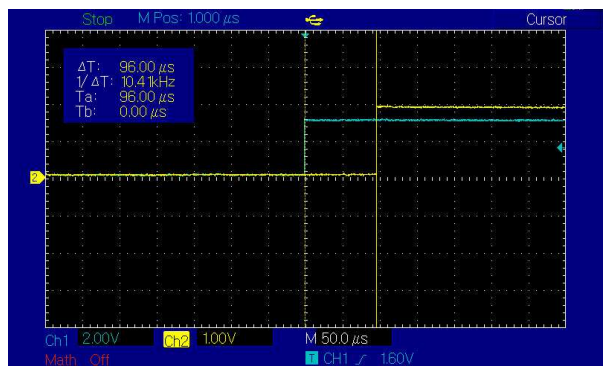

**Figure 10.** STM32F4 Discovery: Outlier measured while running SenseVM implementation of button-blinky.

an outlier. The occurrence of these anomalies is likely to be caused by the garbage collector performing a mark phase. There is also a chance that if a large part of the heap is in use, the allocation operation becomes expensive as it may need to search a larger space before finding a free cell using the lazy sweep algorithm.

The figure below shows a response time of 96$\mu$s compared to the approximately 38$\mu$s response time found in most cases.

The worst outlier found so far was measured to 106.8$\mu$s but these seem to be rare and most outliers fell somewhat short of 96$\mu$s. As future work better measurement methodology will be developed and applied. Preferably such a setup should allow scripting of a large number of automated tests.

## References

[1] 2016. *Zephyr OS.* https://www.zephyrproject.org/

[2] 2019. *WAMR - WebAssembly Micro Runtime*. https://github.com/bytecodealliance/wasm-micro-runtime

[3] Thomas W Barr and Scott Rixner. 2014. Medusa: Managing concurrency and communication in embedded systems. In *2014 USENIX Annual Technical Conference (USENIX ATC 14)*. 439–450.

[4] Robert Cooper and Clifford Krumvieda. 1992. Distributed programming with asynchronous ordered channels in distributed ML. In *ACM SIGPLAN Workshop on ML and Its Applications*. 134–150.

[5] Guy Cousineau, Pierre-Louis Curien, and Michel Mauny. 1985. The Categorical Abstract Machine. In *Functional Programming Languages and Computer Architecture, FPCA 1985, Nancy, France, September 16-19, 1985, Proceedings (Lecture Notes in Computer Science, Vol. 201)*, Jean-Pierre Jouannaud (Ed.). Springer, 50–64. https://doi.org/10.1007/3-540-15975-4_29

[6] Conal Elliott and Paul Hudak. 1997. Functional Reactive Animation. In *Proceedings of the 1997 ACM SIGPLAN International Conference on Functional Programming (ICFP '97), Amsterdam, The Netherlands, June 9-11, 1997*, Simon L. Peyton Jones, Mads Tofte, and A. Michael Berman (Eds.). ACM, 263–273. https://doi.org/10.1145/258948.258973

[7] Zephyr examples. 2021. *Zephyr Button Blinky*. https://gist.github.com/Abhiroop/d83755d7a5703f704fbfb9c3d116d87c

[8] Emden R Gansner and John H Reppy. 1991. eXene. In *Third International Workshop on Standard ML, Pittsburgh, PA*.

[9] Damien George. 2014. *Micropython*. https://micropython.org/

[10] Carl Hewitt. 2010. Actor model of computation: scalable robust information systems. *arXiv preprint arXiv:1008.1459* (2010).

[11] Ralf Hinze. 1993. *The Categorical Abstract Machine: Basics and Enhancements*. Technical Report. University of Bonn.

[12] Charles Antony Richard Hoare. 1978. Communicating Sequential Processes. *Commun. ACM* 21, 8 (1978), 666–677. https://doi.org/10.1145/359576.359585

[13] R John M Hughes. 1982. A semi-incremental garbage collection algorithm. *Software: Practice and Experience* 12, 11 (1982), 1081–1082.

[14] Simon Peyton Jones. 2003. *Haskell 98 language and libraries: the revised report*. Cambridge University Press.

[15] Donald Ervin Knuth. 1997. *The art of computer programming*. Vol. 3. Pearson Education.

[16] Xavier Leroy. 1997. The Caml Light system release 0.74. *URL: http://caml. inria. fr* (1997).

[17] Henry Lieberman and Carl Hewitt. 1983. A Real-Time Garbage Collector Based on the Lifetimes of Objects. *Commun. ACM* 26, 6 (1983), 419–429. https://doi.org/10.1145/358141.358147

[18] Tommi Mikkonen and Antero Taivalsaari. 2008. Web Applications - Spaghetti Code for the 21st Century. In *Proceedings of the 6th ACIS International Conference on Software Engineering Research, Management and Applications, SERA 2008, 20-22 August 2008, Prague, Czech Republic*, Walter Dosch, Roger Y. Lee, Petr Tuma, and Thierry Coupaye (Eds.). IEEE Computer Society, 319–328. https://doi.org/10.1109/SERA.2008.16

[19] Norman Ramsey. 1990. *Concurrent programming in ML*. Technical Report. Princeton University.

[20] John H Reppy. 1992. *Higher-order concurrency*. Technical Report. Cornell University.

[21] John H Reppy. 1993. Concurrent ML: Design, application and semantics. In *Functional Programming, Concurrency, Simulation and Automated Reasoning*. Springer, 165–198.

[22] Abhiroop Sarkar and Mary Sheeran. 2020. Hailstorm: A Statically-Typed, Purely Functional Language for IoT Applications. In *PPDP '20: 22nd International Symposium on Principles and Practice of Declarative Programming, Bologna, Italy, 9-10 September, 2020*. ACM, 12:1–12:16. https://doi.org/10.1145/3414080.3414092

[23] Herbert Schorr and William M. Waite. 1967. An efficient machine-independent procedure for garbage collection in various list structures. *Commun. ACM* 10, 8 (1967), 501–506. https://doi.org/10.1145/363534.363554

[24] Fridtjof Siebert. 1999. Hard Real-Time Garbage-Collection in the Jamaica Virtual Machine. In *6th International Workshop on Real-Time Computing and Applications Symposium (RTCSA '99), 13-16 December 1999, Hong Kong, China*. IEEE Computer Society, 96–102. https://doi.org/10.1109/RTCSA.1999.811198

[25] Robbert Gurdeep Singh and Christophe Scholliers. 2019. WARDuino: a dynamic WebAssembly virtual machine for programming microcontrollers. In *Proceedings of the 16th ACM SIGPLAN International Conference on Managed Programming Languages and Runtimes, MPLR 2019, Athens, Greece, October 21-22, 2019*, Antony L. Hosking and Irene Finocchi (Eds.). ACM, 27–36. https://doi.org/10.1145/3357390.3361029

[26] VDC Research Survey. 2011. *Embedded Engineer Survey Results*. https://blog.vdcresearch.com/embedded_sw/2011/06/2011-embedded-engineer-survey-results-programming-languages-used-to-develop-software.html

[27] Mads Tofte and Jean-Pierre Talpin. 1997. Region-based Memory Management. *Inf. Comput.* 132, 2 (1997), 109–176. https://doi.org/10.1006/inco.1996.2613

[28] Nicolas Tsiftes and Thiemo Voigt. 2018. Velox VM: A safe execution environment for resource-constrained IoT applications. *J. Netw. Comput. Appl.* 118 (2018), 61–73. https://doi.org/10.1016/j.jnca.2018.06.001

[29] Steven Varoumas, Benoît Vaugon, and Emmanuel Chailloux. 2016. Concurrent Programming of Microcontrollers, a Virtual Machine Approach. In *8th European Congress on Embedded Real Time Software and Systems (ERTS 2016)*. 711–720.

[30] Steven Varoumas, Benoît Vaugon, and Emmanuel Chailloux. 2018. A Generic Virtual Machine Approach for Programming Microcontrollers: the OMicroB Project. In *9th European Congress on Embedded Real Time Software and Systems (ERTS 2018)*.

[31] Gordon Williams. 2012. *Espruino*. http://www.espruino.com/

[32] Zephyr. 2016. *Zephyr Mutex API*. https://docs.zephyrproject.org/apidoc/latest/group__mutex__apis.html

[33] Zephyr. 2016. *Zephyr Semaphore API*. https://docs.zephyrproject.org/apidoc/latest/group__semaphore__apis.html

[34] Zephyr. 2016. *Zephyr Semaphore API*. https://docs.zephyrproject.org/apidoc/latest/group__thread__apis.html